\documentclass[a4paper,aps,prb,showpacs,twocolumn]{revtex4}
\usepackage{epsfig}
\usepackage{amsmath}
\begin{document}
\title{Impurity and band effects competition on the appearence of Inverse Giant 
Magnetoresistance in Cu/Fe multilayers with Cr}
\date{\today}
\author{J. Milano} \email{milano@cnea.gov.ar}
\author{A.M. Llois}
\affiliation{Departamento de F\'{\i}sica, Comisi\'on Nacional de Energ\'{\i}a At\'omica. Av. Gral. Paz 1499 - 
(1650) San Mart\'{\i}n, Argentina.}
\affiliation{Departamento de F\'{\i}sica \textquotedblleft Juan Jos\'e Giambiagi\textquotedblright, Facultad de
Ciencias Exactas y Naturales, Universidad de Buenos Aires. Pabell\'on I,
Ciudad Universitaria - (1429) Buenos Aires, Argentina.}
\author{L.B. Steren}
\affiliation{Centro At\'omico Bariloche and Instituto Balseiro. (8400) S.C. de Bariloche,
Argentina}
\begin{abstract}
We have studied the dependence of impurity \textit{vs.} band effects in the
appearance of inverse giant magnetoresistance (IGMR) in Cu/Fe 
superlattices whit Cr.
Current in plane (CIP) and current perpendicular to the plane 
(CPP) geometries are considered.
For the calculation of the conductivities we have used the
linearized Boltzmann equation in the relaxation time approximation. 
Cr impurity effects are taken into account through the spin dependent 
relaxation times and the band effects through the semiclassical velocities obtained
from the LDA calculated electronic structure.
 The larger the Cr/Fe hybridization strength, the bigger is the tendency towards
 IGMR. In particular, in CIP geometry roughness at these interfaces increases 
the IGMR range. 
 The results are compared with experiments and we conclude that
 the experimental GMR curves can only be explained if Cr bands are present.
\end{abstract}
\maketitle

\section{Introduction}

The fast development of new materials and their applications to nanodevices have
made of transport properties a hot area of research in the last years. Many of these 
nanodevices are based on magnetic multilayers (MMLs), which show the Giant
Magnetoresistance effect (GMR) discovered in 1988 \cite{baibich}, that has produced
a  big impact because of the novel physics responsible of the mechanisms involved in
this phenomenon.

Due to the nanolength scales that are reached by new
devices, transport may show up in two possible regimes: diffusive for length
scales larger than the mean free path, or ballistic if they are shorter.
Actually, in a real MML there could certainly exist an interplay
between both regimes.

The spin-dependent potential seen by the electrons in these nanosystems is responsible
for GMR and this potential can be classified as being of
two types. One of them is the so-called extrinsic potential given by scattering with
defects in the bulk and at interfaces and it is usually assumed to be the main
source of
GMR in the diffusive regime. The other type of spin-dependent potential, the
intrinsic one, is determined by potential steps at the ideal
interfaces of the MMLs. If the system is
considered periodic, all information about the intrinsic potential is given
by the
energy bands. Schep {\it et al.} in Ref. \onlinecite{schep} show that GMR in the
ballistic regime is produced by the intrinsic potential. However, calculations
done within the semiclassical approach, using the Boltzmann equation in the
relaxation time approximation, have accounted for several experimental
results in the diffusive regime\cite{ana,ricky,mertigPRL95}.

The experiments on magnetotransport in
MMLs are mostly performed with the electric current flowing parallel to the interfaces
 (CIP
geometry) because the experimental setup in this geometry is easier to achieve than
the one  corresponding to the current flowing perpendicular to the interfaces (CPP).
But the CPP transport configuration is theoretically easier to understand and to
modelise \cite{gijs}. In CIP geometry the electric transport is always diffusive 
because in this geometry the lenghts travelled by the electrons are much larger than
the mean free path. In CPP it is possible to obtain length scales larger, of the
same order, or shorter than the mean free path. Thereafter in this last geometry,
diffusive, ballistic or an interplay between these regimes can be present or
achievable. But, questions concerning the
relative importance of different factors on GMR such as (a) interfacial roughness
and/or interdiffusion \cite{schullerbilayer},(b) competition among the
different length scales \cite{bozec} and of (c) bulk \textit{vs.} interfacial
scattering\cite{schullerinterfacially,vouille} are still open and are being actively
investigated.

One of the most successful and frequently used transport models for
MMLs is the Valet-Fert's \cite{valetfert} one (V-F). This simple model is specially
well suited for diffusive transport in CPP configuration. The main idea behind this
description is that electric transport in a MML can be modelised assuming that there
are two currents, a minority and a majority one contributing both independently to
the
total current, and also that each layer of a MML is thick enough to be considered
as a resistor. The MMLs can be regarded as being built by
resistors arranged in series, each resistor being a source of scattering (in the bulk
or at interfaces). For this assumption to be valid there should be no quantum
interference among sources of scattering. If the distance between interfaces is
shorter than the mean free path there will be quantum coherence and this model
breaks down (Bozec {\it et al.}\cite{bozec}).
Actually the V-F  model has been successfully applied even on systems that are far
beyond
the formal validity limits. But, nowadays it is possible to grow MMLs composed of
very thin
layers and due to this fact the different components of these MMLs cannot be treated
as independent resistors and should be treated as a whole
\cite{ana,ricky,mertigPRL95}.

In CPP the characteristic length for transport is the spin diffusion one. In
general, the spin diffusion length is larger than the mean free path or coherence
length. Coherence lengths in MMLs are of the order of layers' thicknesses
due to roughness and spin accumulation at interfaces. In CIP there is no
spin accumulation and the electrons traverse a fewer number of interfaces than in CPP,
this means that if the thickness of the layers in the multilayers  is less than
the mean free path (the coherence limit) a transport model based on band structure
calculations
becomes realistic in the dilute impurity limit.

Experimentally two kinds of GMR can be observed:
direct and inverse. In the first case the resistance decreases and in the second one
it increases with the applied magnetic field. Direct GMR (DGMR) is more commonly
observed than the inverse one (IGMR), which has only been measured in a few
systems\cite{laura,renard,rahmouni}. In particular, George {\it et al.}\cite{laura}
found a small IGMR ratio for transport in the CIP configuration for a Cu/Fe 
multilayer system when a thin Cr layer is intercalated in half of the Fe layers.
This
inverse effect has been attributed to the existence of alternating spin asymmetries
in the samples, due to different coefficients of the spin-dependent scattering of
electrons at the different interfaces of the superlattice\cite{fert1976}
 (this explanation is actually better suited for measurements done in CPP
 geometry). In this experiment, a mixing of low field IGMR, which adds
 to the high field normal DGMR to be attributed to Cr/Fe, is observed. At low field
 ($<$150 G) the IGMR is due to the alternating spin asymmetries of the two different 
 kinds of magnetic layers present in the sample. For larger fields, the normal and
 large DGMR usually observed in Cr/Fe multilayers overweights the low field effect 
 until
 for very large fields the Fe spins within the Fe/Cr/Fe trilayers finally
 align.
 
In this contribution we want to investigate the differences
between the two kinds of transport geometries and also to get insight into the
relative
importance of band and impurity effects in the determination of the GMR ratio, in
particular for multilayer systems of the type (Fe/Cu/Fe/Cr/Fe/Cu)$_{N}$, which we
consider are ideal for this study. We mainly calculate low field GMR ratios,
that is the GMR ratios which correspond to saturation fields for the relatively
weak AF coupling in Cu/Fe multilayers \cite{petroff}.

The questions we want to address are: (a) the relative importance of intrinsic (bands)
\textit{vs.} extrinsic potential (impurities) effects on the observed IGMR in Cu/Fe 
MMLs containing Cr,
(b) the dependence of the GMR
ratio on the number of Cr/Fe interfaces and/or on  the roughness of the
interfaces, and (c) the possible coexistance of IGMR in one geometry and DGMR in the
other. To carry out this study we consider Cu/Fe multilayers in
which one or more layers of Cr atoms are intercalated in the middle of alternating Fe
layers. As Fe and Cr have nearly the same atomic volume one expects Cr to partially
interdiffuse. We also think that depending on growth conditions
interdiffusion is not necessarily complete; it should, thus, be possible to
consider that, in the average, a continuous Cr layer survives leading the Cr band effects 
and that interdiffused
atoms can be regarded as impurities. We compare, then, results of GMR calculations 
for different interfacial arrangements of Fe and Cr atoms in the above
mentioned MMLs and analyse the different contributions leading to IGMR in these
MMLs.

The band
contribution on transport is explicitly taken into account through the Boltzmann
semiclassical
approach within the relaxation time approximation. The relaxation time value is
considered to be due to impurity scattering at interfaces and in the bulk. The goal
is to \textquotedblleft measure\textquotedblright the relative importance of having 
ordered Cr/Fe interfaces \textit{vs.} Cr
impurities
on the sign of the GMR of these systems. The influence of the number of Cr/Fe
interfaces is investigated as well as the effect of having an
ordered alloy (roughness) at the interfaces. The GMR is calculated as a function of 
the relative
concentration of Cu and Cr scatterers which constitute the impurity effect.
This paper is organized as follows: After
this introduction the method of calculation is outlined in Sec. II, while the
results are provided in Sec. III. and concluding in Sec. IV.
\section{method of calculation}
The electronic structure of  the considered
superlattices is obtained
using an all-electron {\it ab initio} method. The calculations are  performed using
the WIEN97 code \cite{blaha}, which is an implementation of the linearized augmented
plane wave method (FP-LAPW), based on Density Functional Theory. The local spin
density approximation (LSDA) for the exchange and correlation energy as given by
Perdew and Wang is used\cite{perdew}.

The conductivities are calculated within the Boltzmann approach in
the relaxation time approximation and no spin-flip scattering is considered. The 
semiclassical 
Boltzmann equation is valid only in the low impurity limit. The conductivity 
tensor is given, then, by\cite{ziman}
\begin{equation}
\sigma_{ij} = \frac{e^2}{8\pi^2} \, \sum_{\nu s}
\tau^{s} \, \int v_{i \nu}^{s} ({\bf k}) \, v_{j \nu}^{s} ( {\bf k} )
\, \delta [\varepsilon_{\nu}^{s} (\, {\bf k} \,)- \varepsilon_F] \, d^3{\bf k},
\end{equation}
$s$ denotes spin index, $\nu$ the band index, $\varepsilon_F$ is
the Fermi energy. The semiclassical velocities $v_{j \nu}^{s}({\bf k})$ are
obtained from the band calculations.  The relaxation time,  $\tau^{s}$, is
\textbf{k} state independent but spin dependent within our model. Following Ref.
\onlinecite{mertigPRL95}, for the determination
of $\tau^{s}$ we assume that the local spin densities of state at the Fermi
level in the magnetic layers of the superlattice have all the same value. We also
assume that there are interdiffused Cu impurities in the Cu/Fe interfaces and Cr
impurities in the Cr/Fe ones as well. It was shown, in Ref. \onlinecite{mertigPRL98},
that magnetic impurities in Cu make a small contribution to the local density of states at the
Fermi level, and should be, thereafter, ineffective for GMR. As the Boltzmann
approximation is valid only in the low impurity limit, we assume that the
concentration of both types of scatterers ($c_{\text{Cr}}$ and $c_{\text{Cu}}$)
is very small.

We are interested in the evolution of GMR as a function of the
relative importance of both types of scatterers through their modification of
$\tau^{s}$. Thus, we assume that for each of the studied
superlattices, the total number of scatterers per unit cell is fixed and it is equal
to a certain constant, $K$, that is
\begin{equation} c_{\text{Cr}}\,N_{\text{Cr/Fe}}+c_{\text{Cu}}\,N_{\text{Cu/Fe}}=K
\qquad c_{\text{Cu}},\:  c_{\text{Cr}}\ll 1, \end{equation}
$N_{\text{A/Fe}}$ is the number of A/Fe interfaces per unit cell (A = Cr or Cu) and
$c_{\text{A}}$ the
atomic concentration of atoms of type A at the corresponding interfaces. We propose
the following expression for the relaxation time averaged over the Fermi surface,
\begin{equation}
\frac{1}{\tau^{s}}=\frac{K}{N} \left(
\frac{1-\bar{x}_{\text{Cr}}}{\tau^{s}_{\text{Cu/Fe}}}+
\frac{\bar{x}_{\text{Cr}}}{\tau^{s}_{\text{Cr/Fe}}} \right),
\end{equation}
where
\[\bar{x}_{\text{Cr}}=\frac{N_{\textrm{Cr/Fe}}\,c_{\textrm{Cr}}}
{N_{\textrm{Cu/Fe}}\,c_{\textrm{Cu}}+N_{\textrm{Cr/Fe}}\,c_{\textrm{Cr}}}\]
and
\[N=N_{\text{Cr/Fe}}+N_{\text{Cu/Fe}}.\]
$\bar{x}_{\text{Cr}}$ is then the Cr
relative scatterer concentration. $\tau_{\text{A/Fe}}$ denotes the relaxation time of Fe
in the presence of A type impurities (A=Cr, Cu). When calculating the GMR ratios the 
factor $K/N$ in Eq. (3) cancels out.

For the
antiparallel configuration the corresponding expression for $\tau^s$, which
mixes local majority and minority relaxation times in subsequent Fe layers has to be
considered.

In our calculations we obtain the values of $\tau^s_{\rm{A/Fe}}$ appearing in expression
(3) from the spin asymmetries, $\beta_{\textrm{A/Fe}}=\tau^{\uparrow}_{\textrm{A/Fe}}/
\tau^{\downarrow}_{\textrm{A/Fe}}$, as given in Ref. \onlinecite{mertigPRL95}, and from 
\begin{equation}
\sigma_{\rm{A/Fe}} =\tau^{\uparrow}_{\textrm{A/Fe}} \tilde \sigma^{\uparrow}_{\textrm{Fe}} + 
\tau^{\downarrow}_{\textrm{A/Fe}} \tilde \sigma^{\downarrow}_{\textrm{Fe}},
\end{equation}
where $\tilde \sigma^{s}_{{\rm Fe}}$ are the isotropic band
contributions to Fe bulk conduction for the corresponding spin
channel divided by its relaxation time. $\tilde \sigma^{s}_{{\rm Fe}}$ 
are obtained from our electronic bands calculations. $\sigma_{\textrm{A/Fe}}$ is the
inverse of the total residual resistivity of bulk Fe in the presence of 1\% A type
impurities, the corresponding values are taken from Ref. \onlinecite{mertigJMMM}. 
See Table I.

We define the GMR coefficient
as
\begin{equation}
{\rm GMR} =\frac{\sigma_{ii}(AP)}{\sigma_{ii}(P)}-1 \qquad
-1<{\rm GMR}< +\infty
\end{equation}
where P (AP) stands for parallel (antiparallel) configuration. If this coefficient is
positive (negative) we are in the presence of IGMR (DGMR). The P configuration
meant in Eq. (5) is the
one corresponding to the Fe layers aligned across Cu (low field saturation). AP
indicates the initial configuration when the Fe magnetic layers separated by Cu are
antiferromagnetically aligned.
\section{results}
The calculations have been done for superlattices grown
along the (001) direction and following the BCC structure of Fe. As the
layers are thin we assume that Cu grows epitaxially on Fe and within the same
structure; this is actually being revealed by x-ray spectroscopy\cite{pizzini}.
The calculations are done for superlattices of the type Fe/Cu/(Fe,Cr)/Cu, using a
varying number of Cu planes and several combinations of planes and atoms in the
(Fe,Cr) region. The in-plane lattice parameter considered is the one corresponding
to the LDA optimised Fe buffer, the interface distances Cu/Fe and Cr/Fe are also 
optimized.
The considered muffin tin radius R$_{mt}$ are
equal to 2.0 atomic units for the three kinds of atoms involved in the studied systems. The cutoff parameter,
that gives the number of plane waves in the interstitial region, is
taken as R$_{mt}$K$_{max}$=8, where K$_{max}$ is the maximum value of the reciprocal
lattice vector used in
the expansion of plane waves in that zone. We find that the optimized interlayer
distance between Fe and Cu layers increases by 5\% with respect to LDA bulk Fe while the
Fe/Cr interfaces relax by about 4\%.

The band structure is calculated using a mesh of 167 \textbf{k} points in the 
full Brillouin zone (FBZ) and the band contribution to $\sigma_{ij}$ in Eq. (1), is obtained 
using a mesh of 20000 \textbf{k} points in FBZ.
In order to obtain the relative importance of band and impurity effects and the
dependence of IGMR on the number of Cr/Fe interfaces on roughness and on
geometry (CIP/CPP), we analyse the following situations: 1. Cr band effects on GMR,
2. Cr and Cu impurity effects on the GMR of Fe$_{N}$/Cu$_{M}$ superlattices,
3. Cr and Cu impurities together with Cr band effects.

\begin{center}
\textbf{1. Cr band effects on GMR.}
\end{center}

In order to obtain the band contribution of Cr on the GMR ratio we do calculations
for a
Fe$_{\text{3}}$/Cu$_{\text{4}}$/Fe/Cr/Fe/Cu$_{\text{4}}$ superlattice as a function
of Cr/Fe
interface distance. We certainly know that the ground state for the number of Cu layers
 considered here is not
AP, and that in fact the maximum AP exchange coupling appears for 8 or 9 layers of Cu.
In spite of this, the
tendencies we are looking for can be drawn from these calculations which are less demanding in
CPU time.

In Fig. 1 we give the results obtained for
CPP and CIP-GMR coefficients. The results given in
this figure do not take into account the variation of $\tau^{s}$ with impurity
concentration, only the band effects on the GMR of this system are considered by
taking $\tau^{s}$ to be the same for both spin channels. For
comparison we also give the GMR values for the superlattice Fe$_{\rm 3}$/Cu$_{\rm
4}$ in Fig. 1(a). Comparing Fig. 1(a) and Fig. 1(b) it can be seen that the
modification of the superlattice bands through the introduction of a Cr monolayer
gives rise to a large variation of the GMR values, specially in CPP geometry. In
Fig. 1(b) the interfacial Cr/Fe distance is equal to the one  corresponding to bulk
Fe. Fig. 1(c) corresponds to a decrease of 6\% in the Cr/Fe interfacial distance,
and Fig. 1(d) to a reduction of  10\% with respect to Fig. 1(b). It can be seen that 
increasing the hybridization between Fe and Cr atoms, the
tendency towards IGMR also increases. The largest tendencies towards IGMR when introducing 
a Cr monolayer and when
increasing the hybridization are clearly
observed in CPP geometry.

\begin{center}
\textbf{2. Cr and Cu impurity effects on Fe$_N$/Cu$_M$ superlattices.}
\end{center}

In this
case Cu and Cr scatterers in the dilute impurity limit are introduced in MMLs
Fe$_{N}$/Cu$_{M}$ (\textit{N}=3, 5 and \textit{M}=4, 8) through the
values of $\tau^{s}$. No Cr band effects are present, that is no continuous or discontinuous Cr layer is
considered,  and only the contribution
of varying the relative Cr scatterer concentration, $\bar{x}_{{\rm Cr}}$, in the low impurity limit on GMR is analysed for
several examples. In Fig. 2 we show the result of changing the number of Cu or Fe
layers as a function of relative Cr {\it vs.} Cu impurity concentration for
superlattices which  do not contain complete or quasicomplete layers of Cr. Only the
tendencies are relevant for this discussion. It is useful to remember that negative values mean
direct GMR. Changing the number of Fe layers does not modify either the tendencies nor
the absolute values of the coefficients. This can be  seen by comparing Fig. 2(a)
and 2(b) where we increase the number of Fe monolayers from three to five in the
superlattices. A tendency towards inverse GMR when increasing $\bar{x}_{\rm Cr}$ can be observed.
CPP-GMR
is more direct than CIP-GMR for almost all values of $\bar{x}_{{\rm Cr}}$,
but within this
approximation both of them remain direct. Increasing the number of Cu  layers
increases the tendency of CIP-GMR towards IGMR and it even goes positive
within a small range  of $\bar{x}_{{\rm Cr}}$ values (see Fig. 2(c)), which is what
is  experimentally
observed for the Cu width considered. The general tendencies remain the same as in the previous cases. The
CPP-GMR ratio
remains direct and almost constant with $\bar{x}_{{\rm Cr}}$ in all the cases studied in
Fig. 2.

In the transport calculations performed for this item the Fermi level, $\varepsilon_{F}$ has been kept
fixed and equal to its self-consistent value in the corresponding
impurity free multilayer. We have made an estimation of the error done when calculating the GMR ratio while
keeping $\varepsilon_{F}$ fixed for the Fe$_3$/Cu$_4$
superlattice.  In this estimation, we consider that the concentration of Cu and Cr impurities in Fe lies
around 5\%, which is a large value for the low impurity limit assumed here.
 For $\bar{x}_{\text{Cr}}$=0.5
the variation in the CPP-GMR ratio due to the Fermi level shift is of 4\%
and of 2\% in CIP, this does not change the observed tendencies.  We consider
that for the other cases here treated the situation is similar
to the present one.

\begin{center}
\textbf{3. Cr and Cu impurity effects and Cr band effects.}
\end{center}

We analyse in the following examples Cr band effects and, simultaneously, Cr and Cu
impurity effects on the GMR ratios of the studied superlattices.
Cr band effects are taken into account  by introducing continuous
and discontinuous Cr layers, and the impurity
effects through the variation of $\tau^{s}$ as a function of relative impurity
concentration in the low impurity limit as in the previous  examples. It is
well known that Cr mixes with Fe, but a certain averaged
ordered Cr configuration should survive after deposition.

In Fig. 3 we show the
effect on the GMR of introducing  a varying number of Cr/Fe interfaces, and also that of
including
more  Cu layers as a function of $\bar{x}_{\rm Cr}$. We also give the values
of GMR when no impurity effects are taken into account (horizontal lines). We see in Fig. 3(a)
that the inclusion of Cr band effects through the presence of one Cr layer in the  unit cell,
drastically modifies CPP-GMR as compared to the results for Fig. 2(a). In the new situation, CPP-GMR
 goes positive for almost all values of
$\bar{x}_{\rm Cr}$. It should be noticed that Cu impurities lower the
GMR coefficient if one compares with the results of calculations that only include band effects
(straight lines in Fig. 3). An increase in $\bar{x}_{{\rm Cr}}$ drives both GMR coefficients positive.

If one changes the number of Cu
layers, CPP-GMR goes down even if it is mostly positive, but
CIP-GMR remains nearly unaffected; see Fig. 3(b). We expect that in the real situation, with
the number of Cu layers lying around 8-9 for the maximum antiferromagnetic coupling,
CPP-GMR should be inverse and larger than CIP-GMR. 
On the other hand, CIP-GMR is nearly not modified when the
number of Cu layers is changed within the widths we are treating in these calculations.

Duplicating the number of Cr/Fe interfaces does not give rise to a significant change in
CIP-GMR while CPP-GMR nearly doubles its value, as it is shown in Fig. 3(c).

We simulate a particular case of
roughness by adding an ordered 50\% Cr and 50\% Fe ML on each side of the Cr layer
(see Fig. 3(d)), and we obtain a larger $\bar{x}_{\rm Cr}$ range for which CIP-GMR is inverse, while
CPP-GMR does not change
with respect to the example of Fig 3(a).  This can be easily understood as the
introduction of roughness  generates Cr/Fe interfaces perpendicular to
the superlattice growth direction, and we have already seen that the presence of
these
interfaces is a source of inverse GMR due to Cr/Fe hybridization. The effect is
nevertheless not large enough as to qualitatively modify the maximum value reached
by the
GMR ratio. CPP-GMR does not change
in this particular example as the number of Cr/Fe interfaces along $z$ is the same as in the 
example of Fig. 3(a). It is interesting to notice that GMR
has a maximum as a function of concentration in each one of the studied cases.

We also simulate the effect of applying large magnetic fields to the samples, 
that is, the process of aligning not only the magnetic moments of Fe across Cu,
but also those of Cr with respect to the adjacent Fe atoms. In our calculations we observe
that Cr has a tendency  to antiferromagnetically align with respect to the
surrounding Fe layers both in the AP configuration as in the P configuration.
To simulate the presence of the external magnetic field, we do fixed spin moment calculations 
by constraining the total
cell magnetic moment and increasing it progressively towards its high field
saturation value. In Table II we show the results obtained  for the example
 Fe$_3$/Cu$_4$/Fe/Cr/Fe/Cu$_4$ and $\bar{x}_{\text{Cr}}$=0.5. We give
the obtained GMR ratios for different values of the total
magnetic moment, together with the local magnetic moments on Cr and on the
neighboring Fe atoms for each constrained total magnetic moment.
The experimental behavior for the evolution of GMR in the presence of growing 
magnetic fields is obtained. That is, an initial increase in the values of the GMR ratios until 
the low saturation
limit is reached and beyond this, a slow decrease in the GMR ratios as a function of growing 
applied magnetic field\cite{laura}. This shows again the importance of band effects on the GMR of the
system under study, as this evolution of GMR would not have been observed if only impurity
effects had been taken into account.
\section{Conclusions}
Based on \textit{ab initio} band structure calculations, we try to determine 
the competition between Cr band and impurity effects
on the type of GMR (direct or inverse) for superlattices of the type
(Fe/Cu/Fe/Cr/Fe/Cu)$_N$. The conductivities used to obtain the GMR ratios are 
calculated 
semiclassically by using the linearized Boltzmann approach. The impurity 
effects are taken into account through an 
averaged relaxation time per
spin channel. We make calculations for, both, CIP and CPP
geometries and work in the low impurity limit. The conclusions drawn can be summarized as follows:

The value and sign of GMR depends strongly on the
hybridization strength between Fe and Cr layers, specially in CPP geometry.

In the absence of Cr band effects and when only isolated Cr and Cu 
impurities in Fe are considered, Fe$_N$/Cu$_M$ superlattices show a clear tendency towards IGMR in 
CIP configuration. This tendency depends on the relative concentration of Cr
\textit{vs.} Cu impurities. For some Cu widths and within a certain
concentration range it even goes positive. In CPP the situation is different, the
GMR ratio is far from being inverse over the whole range of relative impurity
concentrations. These
results had already been observed by P. Zahn \textit{et al.}\cite{mertigPRL95}.
A change in the
number of Fe layers does not modify these tendencies.

When complete or incomplete Cr layers are introduced in alternating Fe
layers and, thereafter, Cr band effects are switched on, quantitative as qualitative
changes take place specially in CPP. Both, in CIP as CPP the GMR ratios acquire a
larger tendency towards IGMR, but in this case it is the CPP geometry the one with
the largest IGMR ratios. In the two geometries the GMR is inverse within a broad range of the Cr and Cu
relative impurity concentration. In CPP, Cr/Fe interface effects are more
important than
impurity ones to determine the type of GMR, while the opposite is true in
CIP. This is being confirmed by the fact that doubling the number of
Fe/Cr interfaces (see Fig 3) gives rise to a large increase in
CPP-IGMR, while the increment in CIP-GMR is not as important.

The introduction of roughness (in our case an ordered roughness) at the 
interfaces,
gives rise to an increasing tendency towards CIP-IGMR. Electrons flowing
in CIP geometry face the appearance of effective Cr/Fe interfaces in the 
presence of roughness and Cr band effects become more relevant in this geometry.

We have shown that the experimentally observed evolution of GMR for Fe$_N$/Cu$_M$ MMLs with Cr, as a function 
of an increasing external magnetic field, can be explained if the presence of Cr bands is assumed.

In general, both transport geometries share  the same  tendencies  for 
 the sign of the GMR ratio,  even if the presence  of Cr band effects 
makes CPP more liable to IGMR than CIP, contrary to what happens when 
only Cr impurities are considered.

Summarizing, both disorder as band effects are 
necessary to explain the appearance and evolution of IGMR in the studied
superlattices.

\begin{acknowledgments}
We thank Dr. M. Alouani  for helpful and fruitful discussions.
Two of us (A.M.L. and L.B.S.) acknowledge Consejo Nacional de Investigaciones
Cient\'{\i}ficas y T\'ecnicas for support of this work.
We acknowledge Fundaci\'on Sauberan, Fundaci\'on Antorchas and Fundaci\'on Balseiro also for support.
This work was partially funded by project UBACyT X115.
\end{acknowledgments}

\newpage
\begin{table}
\begin{ruledtabular}
\begin{tabular}{ccc}
                                 & \multicolumn{2}{c}{Impurity (A)} \\ \cline{2-3}
                                 & Cr   & Cu                    \\ \hline \hline
$\sigma^{\uparrow}_{\rm{A/Fe}}   $  & 0.12  & 0.53              \\ \hline
$\sigma^{\downarrow}_{\rm{A/Fe}} $  & 0.70   & 0.065            \\ \hline
 $\beta_{\rm{A/Fe}}$             & 0.11 & 3.68                  \\ \hline
 $\tau^{\uparrow}_{\rm{A/Fe}}$   & 1.05 & 4.84                  \\ \hline
 $\tau^{\downarrow}_{\rm{A/Fe}}$ & 9.75 & 1.31                  \\
\end{tabular}
\end{ruledtabular}
\caption{$\sigma^s_{\rm{A/Fe}}$ is the inverse of the residual resistivity of Fe bulk in the
presence of A-type impurities (1\%); the values are
taken from Ref. \onlinecite{mertigJMMM} and are given in $(\mu\Omega.cm)^{-1}$. $\beta_{\rm{A/Fe}}$ 
means the asymmetry coefficient of
Fe in the presence of A-type impurities; the values are taken from Ref. \onlinecite{mertigPRL95}.
$\tau^s_{\rm{A/Fe}}$ are the Fe relaxation times per spin channel obtained using Eq. (4) and are given in
arbitrary units.}
\label{Table I}
\end{table}
\begin{table}
\begin{ruledtabular}
\begin{tabular}{cccccc}
 Magnetic\\ configuration  & $\mu_{\textrm{cell}}$  & $\mu_{Fe}$   &$\mu_{Cr}$&  CIP-GMR & CPP-GMR   \\ \hline \hline
 AP             & -4.88 & 1.14 & -0.22                & 0    & 0     \\ \hline
 P              & 8.82 & 1.02 & -0.25                & 0.12 & 0.65  \\ \hline
 FSM1		& 13.00 & 1.95 & 0.30                 & -0.20& 0.27  \\ \hline
 FSM2		& 16.00 & 2.42 & 0.98                 & -0.30& -0.34 \\
\end{tabular}
\end{ruledtabular}
\caption{CIP and CPP-GMR coefficients calculated for different initial magnetic configurations of the
superlattice
Fe$_3$/Cu$_4$/Fe/Cr/Fe/Cu$_4$. $\mu_{\rm{cell}}$ denotes the total magnetic moment per unit
cell and $\mu_{\rm{A}}$ the local
magnetic moments on Cr or on the Fe atoms adjacent to the Cr layer. AP denotes the initial zero
field magnetic configuration, P
means the low field saturation configuration (Fe layers aligned across Cu), FSM1 and FSM2
stand for configurations in which
the total magnetic moment is larger than for the P one. All moments are given in units of
$\mu_{\rm{B}}$.}
\label{Table II}
\end{table}
\clearpage

\begin{figure}[t]
\epsfysize=1.8in
\epsfbox{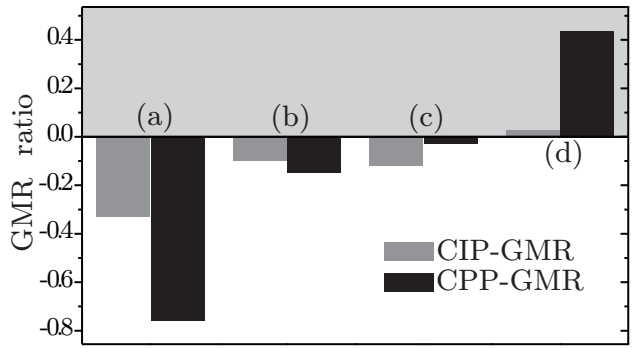}
\caption{Calculated GMR for a) Fe$_3$/Cu$_4$, b) Fe$_3$/Cu$_4$/Fe/Cr/Fe/Cu$_4$ with interfacial Cr/Fe
distance equal to
the one of bulk Fe, c) and d) idem b) but with Cr/Fe
distance 6\%  and 10\% respectively smaller. The superlattices are grown along the (001) direction. In this
case the in plane lattice parameter are those of Fe bulk.}
\end{figure}

\begin{figure}
\epsfysize=4.5in
\epsfbox{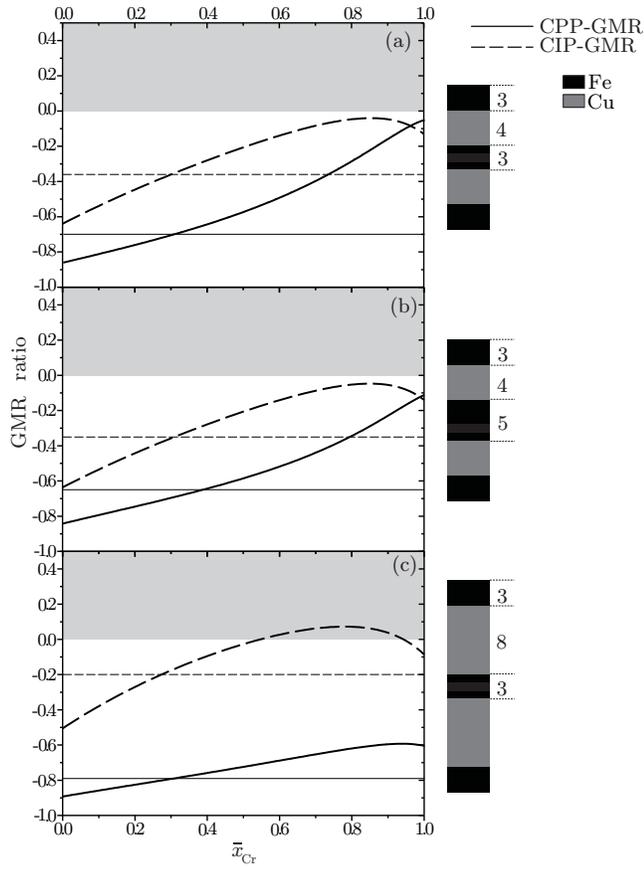}
\caption{Calculated GMR for a)
Fe$_3$/Cu$_4$, b) Fe$_5$/Cu$_4$ and c) Fe$_3$/Cu$_8$ as a function of the relative Cr
scatterer concentration, $\bar{x}_{\textrm{Cr}}$.
Shadowed areas correspond to IGMR. Straight lines give the GMR ratios in the absence of 
impurity scatterers. No Cr band effects present. The number of atomic layers in each layer of the MMLs is
given on the right of the schematic MMLs.}
\end{figure}

\begin{figure}
\epsfysize=5.5in
\epsfbox{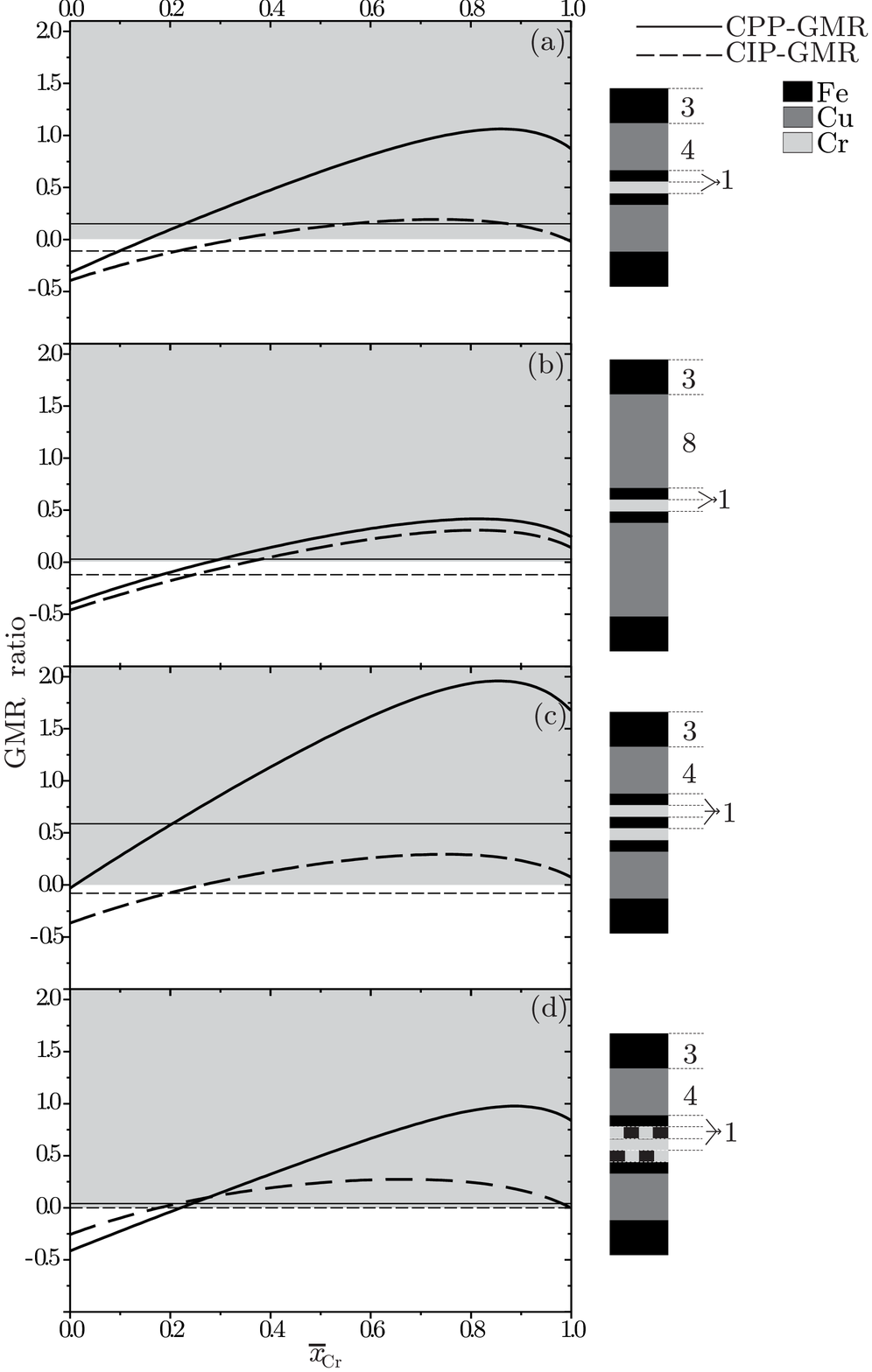}
\caption{Calculated GMR for a) Fe$_3$/Cu$_4$/Fe/Cr/Fe/Cu$_4$, b)
Fe$_3$/Cu$_8$/Fe/Cr/Fe/Cu$_8$, c) Fe$_3$/Cu$_4$/Fe/Cr/Fe/Cr/Fe/Cu$_4$, d)
Fe$_3$/Cu$_4$/Fe/Fe$_{0.5}$Cr$_{0.5}$/Fe/Fe$_{0.5}$Cr$_{0.5}$/Fe/Cu$_4$ as a function of the
relative Cr/Cu scatterer
concentration, $\bar{x}_{\textrm{Cr}}$. Straight lines give the GMR ratios in the absence of impurity
scatterers. The number of atomic layers in each layer of the MMLs is
given on the right of the schematic MMLs.}
\end{figure}

\begin{thebibliography}{999}
\bibitem{baibich}
M.N. Baibich, J.M. Broto, A. Fert, F. Nguyen Van Dau, F. Petroff, P.
Etienne, G. Creuzet, A. Friederich, J. Chazelas, Phys. Rev. Lett. \textbf{61},
 2472 (1988).
\bibitem{schep}
K.M. Schep, P.J. Kelly, G.E.W. Bauer, Phys. Rev. Lett. \textbf{74}, 586 (1995).
\bibitem{ana}
M. Weissmann, A.M. Llois, R. Ram\'{\i}rez, M. Kiwi, Phys. Rev. B \textbf{54}, 15335
 (1996).
\bibitem{ricky}
R. G\'omez Abal, A.M. Llois, M. Weissmann, Phys. Rev. B \textbf{53}, R8844 (1996).
\bibitem{mertigPRL95} P. Zahn, I. Mertig, M. Richter, and H. Eschrig, Phys. Rev.
 Lett. \textbf{75}, 2996 (1995).
\bibitem{gijs}
M.A.M. Gijs, G.E.W. Bauer, Adv. Phys. \textbf{46}, 285 (1997).
\bibitem{schullerbilayer}
M.C. Cyrille, S. Kim, M.E. Gomez, J. Santamaria, K.M. Krishnan, I.K. Schuller, Phys. Rev. B
 \textbf{62}, 3361 (2000).
\bibitem{bozec}
D. Bozec, M.A. Howson, B.J. Hickey, S. Shatz, N. Wiser, E.Y. Tsymbal, D.G. Pettifor, Phys. Rev. Lett.
\textbf{85}, 1314 (2000).
\bibitem{schullerinterfacially}
J. Santamaria, M.E. Gomez, M.C. Cyrille, C. Leighton, K.M. Krishnan, I.K. Schuller, Phys. Rev. B \textbf{65},
012412-1 (2001).
\bibitem{vouille}
C. Vouille, A. Barth\'el\'emy, F. Elokan Mpondo, A. Fert, P.A. Schroeder, S.Y. Hsu, A. Reilly, R. Loloee,
Phys. Rev. B \textbf{60}, 6710 (1999).
\bibitem{valetfert} T. Valet, and A. Fert, Phys. Rev. B \textbf{48}, 7099 (1993).
\bibitem{laura}
J.M. George, L.G. Pereira, A. Barth\'el\'emy, F. Petroff, L. Steren, J.L. Duvail, A. Fert, R. Loloee,
P. Holody, P.A. Schroeder, Phys. Rev. Lett. \textbf{72}, 408 (1994).
\bibitem{renard}
J.-P. Renard, P. Bruno, R. M\'egy, B. Bartenlian, P. Beauvillain, C. Chappert, C. Dupas, E. Kolb, M. Mulloy,
P. Veillet, E. V\'elu, Phys. Rev. B \textbf{51}, 12821 (1995).
\bibitem{rahmouni}
K. Rahmouni, A. Dinia, D. Stoeffler, K. Ounadjela, H.A.M. Van den Berg, H. Rakoto, Phys. Rev. B \textbf{59},
9475 (1999).
\bibitem{fert1976}
A. Fert, I.A. Campbell, J. Phys. F \textbf{6}, 849 (1976).
\bibitem{petroff}
F. Petroff, A. Barth\'el\'emy, D.H. Mosca, D.K. Lottis, A. Fert, P.A. Schroeder, W.P. Pratt, Jr., R. Loloee, S.
Lequien, Phys. Rev B \textbf{44},  5355, (1991).
\bibitem{blaha}
P. Blaha, K. Schwarz, J. Luitz, WIEN97, Vienna University of
Technology, Vienna 1997. (Improved and updated Unix version of the original
copyrighted WIEN-code, which was published by P. Blaha, K. Schwarz,
P. Sorantin, S. B. Trickey, in Comput. Phys. Commun. \textbf{59}, 399 (1990).
\bibitem{perdew}
J.P. Perdew, Y. Wang, Phys. Rev. B \textbf{45}, 13244 (1992).
\bibitem{ziman}
J.M. Ziman, \textit{Electrons and Phonons} (Oxford University Press, London, 1967), Chap. VII.
\bibitem{mertigPRL98} P. Zahn, J. Binder, I. Mertig, R. Zeller, P.H. Dederichs, Phys. Rev.
 Lett. \textbf{80}, 4309 (1998).
\bibitem{mertigJMMM} I. Mertig, P. Zahn, M. Richter, H. Eschrig, R. Zeller, and P.H.
 Dederichs, J. Magn. Magn. Mater. \textbf{151}, 363 (1995).
\bibitem{pizzini}
S. Pizzini, F. Baudelet, D. Chandesris, A. Fontaine, H. Magnan, J.M. George, F. Petroff, A. Barth\'el\'emy,
A. Fert, R. Loloee, P.A. Schroeder, Phys. Rev. B \textbf{46}, 1253 (1992).
\end{thebibliography}
\end{document}